# NiCE Teacher Workshop: Engaging K-12 Teachers in the Development of Curricular Materials That Utilize Complex Networks Concepts

Emma K. Towlson[1], Lori Sheetz[2], Ralucca Gera[3], Jon Roginski[4], Catherine Cramer[5], Stephen Uzzo[6], and Hiroki Sayama[7]

1. Center for Complex Network Research, Northeastern University, USA
2. Center for Leadership and Diversity in STEM, U.S. Military Academy at West Point, USA
3. Department of Applied Mathematics, Naval Postgraduate School, USA
4. Network Science Center, U.S. Military Academy at West Point, USA
5. Data Science Institute, Columbia University
6. New York Hall of Science, USA
7. Center for Collective Dynamics of Complex Systems, Binghamton University, USA

Corresponding author: ektowlson@gmail.com

**Abstract**

Our educational systems must prepare students for an increasingly interconnected future, and teachers require equipping with modern tools, such as network science, to achieve this. We held a Networks in Classroom Education (NiCE) workshop for a group of 21 K-12 teachers with various disciplinary backgrounds. The explicit aim of this was to introduce them to concepts in network science, show them how these concepts can be utilized in the classroom, and empower them to develop resources, in the form of lesson plans, for themselves and the wider community. Here we detail the nature of the workshop and present its outcomes – including an innovative set of publicly available lesson plans. We discuss the future for successful integration of network science in K-12 education, and the importance of inspiring and enabling our teachers.

**Introduction, background, and motivation**

We live in a highly interconnected world, in which networks are ever more important to understand if we wish to successfully navigate it (1). Yet, our education systems do not adequately reflect this modern reality, instead remaining surprisingly modular, with subjects taught in an isolated, linear fashion. Network concepts have the potential for broad impacts, as a way of thinking, a toolset, and a style of teaching that facilitates cross-disciplinary learning in a manner suited to the connected way in which children think (2).

In previous projects to bring network science and network-thinking to high schools (3–5), we identified a massive need for accessible networks related educational resources. Even for the graduate and undergraduate level student, such resources are new and few in number (6). This motivated the production of a Networks Literacy: Essential Concepts and Core Ideas booklet (7,8), which forms the crux of a basic networks understanding and is now available in 20 languages.

To create and disseminate the highest quality network science related resources for and to schools, there is clearly a need to incorporate two distinct skillsets: that of the practicing network scientist, and that of the practicing teacher. The network scientist brings the subject-specific knowledge and expertise to accurately convey the new concepts, and make full use of the available toolsets. The teacher brings the wealth of classroom experience and knowledge of formal and effective lesson design. Both parties are of the utmost importance in resource creation, and, ideally, the lead should be taken by the practitioner: the teacher.

We recognized that the shift in thinking associated with network science is something that teachers need support for. When learning alongside their students, the comfort level of teachers was very low: they are accustomed to being the authority in the classroom. We therefore identified a need to create a space for teachers to learn, and gain ownership of the initiative of bringing network science and network-thinking to their schools. This led to the conception of the NiCE workshop.

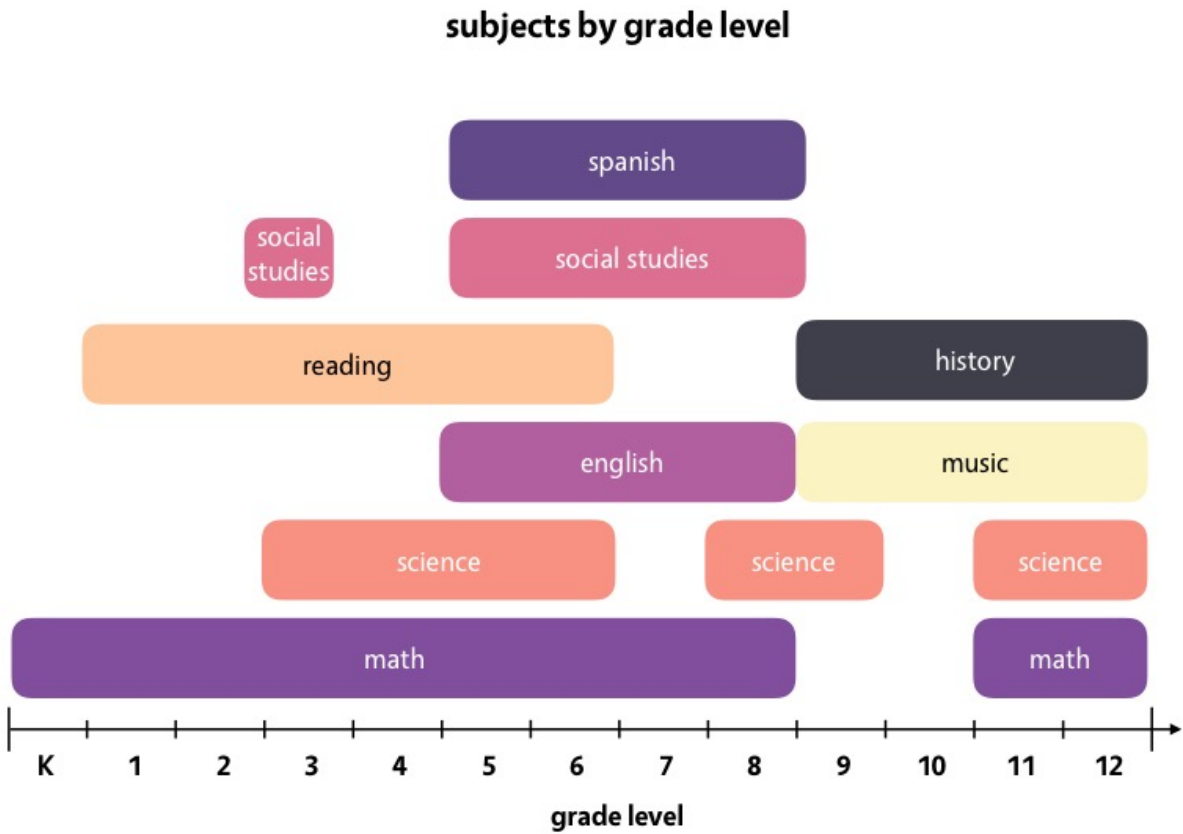

**Figure 1**: **Coverage of grades and subjects by skillsets of workshop participants.** Educators who attended the workshop have highly diverse skillsets, both in terms of age/grade level of student, and specialist subjects. The whole K-12 grade levels were successfully spanned, with good representation in most curricular areas. Note also that one teacher specializes in professional development, and another in intervention; they are not displayed in the figure.

**Concept**

A core principle of our vision is that network science should be (a) pervasive as a way of thinking regardless of subject area; and (b) something which *helps* teachers with their existing tasks (9). A teacher's job is extremely demanding, and necessarily involves meeting a number of governmental standards. With the introduction of the Next Generation Science Standards (NGSS) arriving in parts of the USA (10), teachers of scientific subjects face the extra challenge of an official move towards connected thinking, in the form of "cross-cutting concepts". (At the time of the workshop, the teachers from the West coast had been subject to these standards for a year and those from the East coast had not.) Network science is then ideally suited to tackling these new requirements.

Resources for teachers are best developed by teachers themselves, and we aim to empower and enable this development. By educating teachers on networks concepts, providing a creative and comfortable environment within which to spark ideas, and providing the support and expertise of practicing network scientists, we set out to foster the ideal first focus group to create such resources. We want to show the education community the ways that network science can be a tool for their teaching, and inspire teachers to take ownership of their own network science related initiatives.

**Execution**

We invited applications from teachers in a number of schools and locations in the regions of the organizers (predominantly on the East and West coasts of the USA). These were chosen to ensure our ability to provide adequate pre- and post-workshop support. From the applications, we selected 21 individuals to attend the workshop. These educators brought more than 300 years of cumulated teaching experience, and a highly diverse set of skills. Their knowledge spanned subjects including math, English, Spanish, science, social studies, history, and music, and represented all grade levels from K-12, plus educational coaches (see Figure 1).

We held the workshop at the United States Military Academy at West Point (USMA) over the course of four days, two full and two partial, in July 2017. Being away from school, short, and in early summer, it was intended to be non-intrusive on school or holiday time. We also provided a small stipend to remunerate teachers for their significant efforts. The schedule included a blend of lectures, discussion groups, and interactive/hands-on sessions – see Table 1 for a summary, and the workshop website for more details (11). For the purpose of effective collaboration, teachers were seated in five groups, each with a different focus area according to their areas of expertise: K-2nd grade, 3-5th grade, 6-8th grade, High School, and Educational Coach. Each group was partnered with a network scientist, from the Center for Complex Network Research (CCNR) at Northeastern University, the USMA Network Science Center, or the New York Hall of Science (NYSCI).

*Day 1:* Before arriving, participants were asked to review three short online videos to introduce the idea of network science (11). The workshop began with a low-key evening intended for introductions, motivating the themes and goals of the workshop, and brainstorming existing conceptions of networks and any questions people had from the pre-workshop material.

| |
|---|
| **Day 1 (evening): Welcome dinner and movie** |
| Social time and dinner |
| Introductions and brainstorming on network concepts |
| Movie: *Connected* |
| **Day 2: Hands on network science and concept development** |
| Introduction to network science (presentation) |
| Spelling and networks (activity) |
| Network Literacy Essential Concepts and Core Ideas (presentations) |
| NGSS and network science (presentation and activity) |
| Brainstorming – networks questions and applications |
| Team work, creating ideas for lesson plans |
| Introduction to *Gephi* – hands-on |
| Spreading on networks (presentation) |
| Developing networks from module ideas (activity) |
| Group speed rounds |
| Tour of West Point |
| **Day 3: Lesson planning, development, and rehearsal** |
| Developing lesson plans |
| Network robustness (activity) |
| Developing network of governing standards of module, refining lesson plans |
| Group speed rounds and feedback |
| Boat ride |
| Refining lesson plans, preparing outputs |
| Group speed rounds and feedback |
| **Day 4 (morning): Group outbriefs** |
| Post workshop survey |
| Group outbriefs |

**Table 1: Schedule overview.** Further details are available on the NiCE workshop website (11).

*Day 2*: The first full day introduced crucial network science concepts and reviewed the structure and standards for modules and lesson plans (with a focus on the NGSS (10)). This was accomplished via presentations, hands-on activities, and group and whole-room discussions. The intention of this day was to lay the groundwork for resource development, by educating about network science and helping the teachers identify key areas which could benefit from the network science lens.

*Day 3:* The second full day was benchmarked for substantial development of lesson plans and projects. It included a hands-on network science activity, and the mapping of standards for curricular modules.

*Day 4*: On the final morning, the teachers presented their materials, and were provided instructions for completing their projects for review; they were asked to do this within two weeks of departure. They were also asked to complete a post-workshop survey, so that we could gather more metrics with which to gauge its success and impact (see Outcomes section below), and information about the best ways to move forwards.

We also included a tour of USMA (on Day 2), and a boat ride (on Day 3), which fostered working relationships, inspiration, and creativity. At multiple stages throughout the workshop, groups were given the opportunity to present their ideas and progress on lesson plans, for feedback from network scientists and the other participants. This allowed everyone to incorporate a variety of suggestions and fully develop their concepts.

**Figure 2**: **Wordcloud of participants' overall experience.** The participating teachers were asked to describe their experience at the NiCE workshop, using only three words. The figure illustrates the words they put forward, with font sized by number of mentions; for reference, "informative" was put forward six times, and "stimulating" one time. Variations of the same concept (e.g. "educational" and "educationally") are agglomerated into one representative word. Responses were overwhelmingly positive, and clearly demonstrate significant learning, and creative sparks for future endeavors.

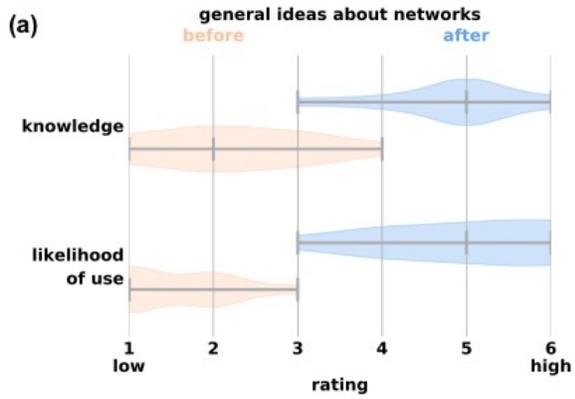
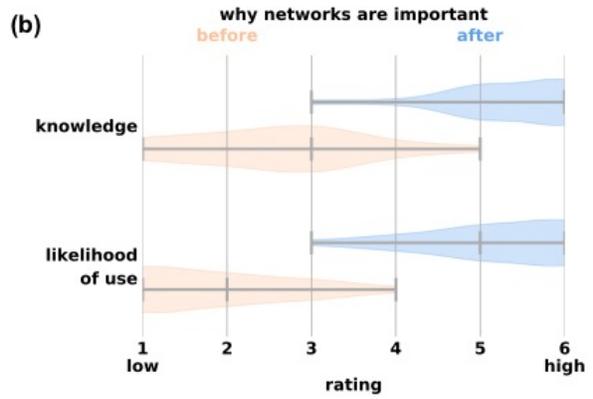
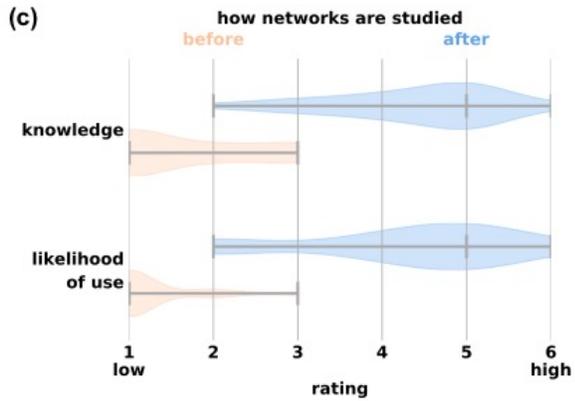
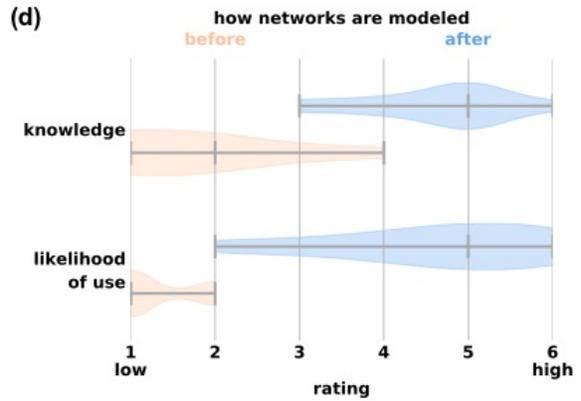
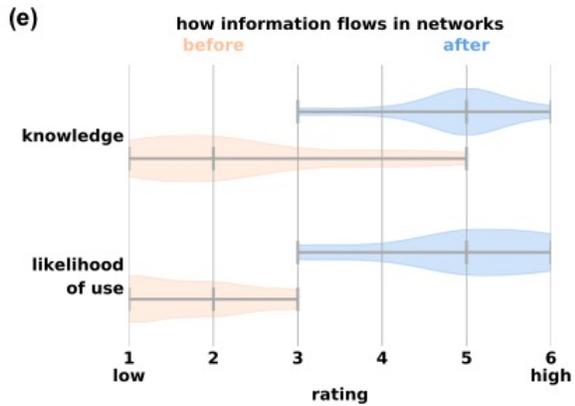
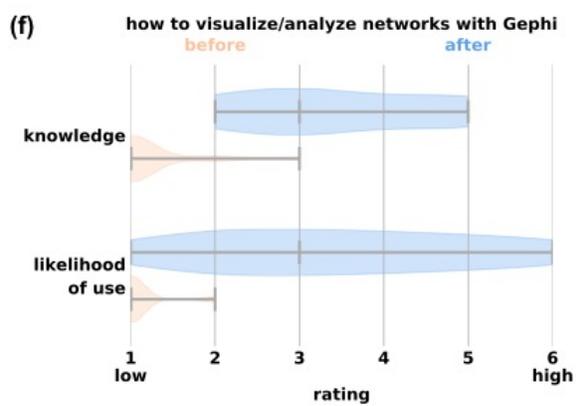
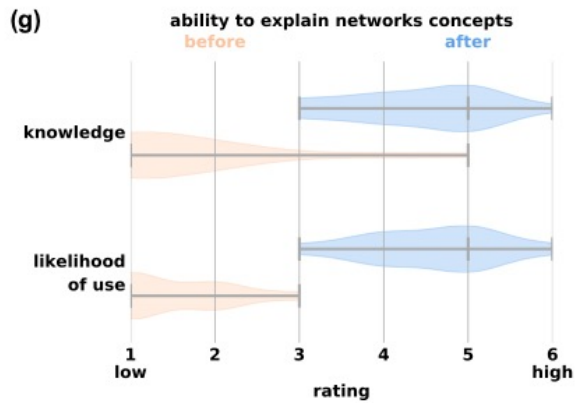

**Figure 3**: **Violin plots illustrating participants' knowledge and use of networks, before and after the workshop.** Following the workshop, the participating teachers were asked to rate a number of network related queries (on a scale of 1-6, with 1 the lowest and 6 the highest). For each query, they were asked to rate (i) their knowledge before the workshop; (ii) their knowledge after the workshop; (iii) their likelihood of using it in the classroom before the workshop; (iv) their likelihood of using it in the classroom after the workshop. On all points, the teachers indicated large increases in both their understanding of the network concept, and in their likelihood to make use of the network concept in their lessons.

**Outcomes: findings and end products**

The workshop was a vibrant, highly creative experience, which spurred lots of discussions about the utility of the lens of networks thinking, both in lessons and in lesson/curriculum planning. In the end of workshop survey, the teachers were asked to describe their experience using only three words. The results are shown in Figure 2, and provide a picture of a motivating and helpful learning experience. The most commonly put forward words include "informative", "inspiring", "relevant", "thoughtful", "connected", and "interesting". They were further queried on a number of network science related points, such as "the importance of networks" (see Figure 3). They were asked to rate their knowledge of each point before the workshop, their knowledge of each point after the workshop, their likelihood to use each point before the workshop, and their likelihood to use each point after the workshop. In all cases, there is a large increase in the before and after ratings, demonstrating the success (i) of conveying the network science concepts; and (ii) in motivating the teachers to use these concepts in their own classrooms.

In all, we received 13 exceptionally innovative resources for the teaching community. They comprise 10 lesson plans, and 3 network science in education projects (a professional development module, a plan to map an effective calculus revision guide, and a plan for a social network analysis to improve management and organization). The resources are summarized in Table 2 and are all available online (11).

For example, one lesson aimed at elementary to middle school students explores the factors underpinning the survival of an ecosystem, using network science to explore the relationships between the data (see Figure 4 (a-b)). Students are invited to examine plants in the school grounds, and note properties such as color, location, and scent. They are challenged to identify cause and effect, and in doing so, earn an integrative understanding of the local ecosystem. Another lesson, designed for high school history students, considers the globalization of trade networks (see Figure 4 (c)) (12). By mapping these networks and considering areas of high traffic (ports, and popular trading routes), one can elucidate the important connections and valuable resources. And indeed, how and why these connections formed, and the relationship with global political power.

**Discussion and outlook**

Successful integration of network science and networks-thinking with high school education necessarily relies on educators. There is currently a lack of material to use in lesson design, presenting a need and an opportunity. The best resources, including lesson plans, will be developed via collaborative efforts between teachers and network scientists, thus bringing together the experience, knowledge, skillsets, and ways of thinking of both types of experts. To make it worthwhile for teachers to invest in such efforts on top of their existing demands, the payoff to them of network science must be made explicit. Lesson content can be created which cuts across concepts and curricular subjects, and satisfies the governmental standards. This facilitates collaboration with other teachers and departments, increases creativity, and eases workload. Further, network science can be a valuable tool in understanding and effectively implementing these standards (in particular, the interconnected criteria in the NGSS).

| Lesson plans | | Level |
|---|---|---|
| Three skeleton key | Reading and comprehension, considering the spread of plague on a network in the Middle Ages. | Middle |
| Depletion of fish in oceans | Comprehensive lesson drawing links between various reading materials and acknowledging the interconnectedness of factors harming the ocean ecosystems. | Middle |
| Balance networks in music ensembles | Identifying balance networks in a string orchestra, and how alterations to the network change resultant sound. | High |
| Network science in relationship development | Understanding and developing personal relationships and strong communities through the lens of network science. | Elementary |
| Interdependent relationships in an ecosystem | Exploring the connected factors at play in the growth and survival of plants in the school grounds. | Elementary & Middle |
| Nanotechnology | Use connections between words to clarify concepts and meanings in a text, and then connections between claims and reasoning to develop an argument. | Middle |
| The power of literate people | Exploring connections between reading, behavior, and literacy. | Elementary |
| Applying network science to resource management | Map the structure of resource reliance on common materials and the relation to waste. | High |
| Shark network | Consequences of the portrayal of sharks in the media. | Middle |
| Networks in world history | Examining trade networks, and the effect of globalization of such networks with the Columbian Exchange. | High |
| **Network science and education projects** | | |
| Networking in a calculus environment | Analyze topics of questions in past AP calculus exams to design effective revision strategies. | High |
| Social network within the school | Map communication networks between staff, to improve organization and management. | Teacher |
| Networking interdisciplinary grade level standards | Professional development module. Generate network maps of standards to identify opportunities for interdisciplinary lessons and curriculum development. | Teacher |

**Table 2: Network science in education resources developed by the teachers at the NiCE workshop.** Titles, brief summaries, and target level are provided. Lesson plans and projects can all be found on the NiCE website (11).

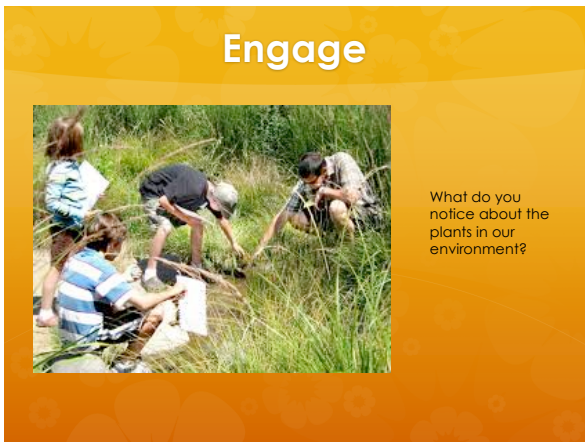
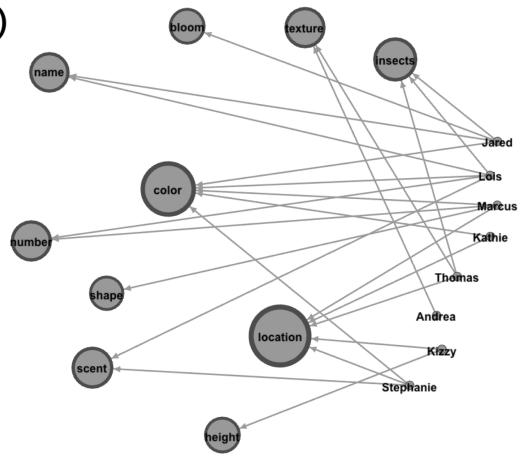
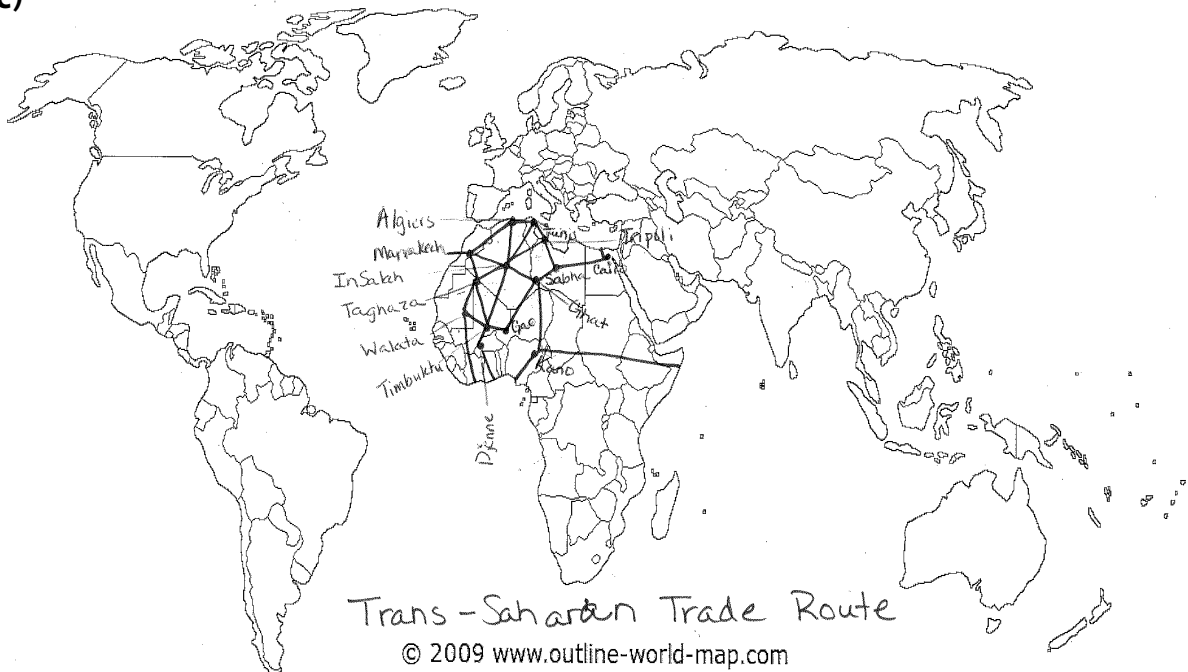

**Figure 4: Excerpts from example lesson plans.** Students are invited to make observations about plants in the school grounds (a), and to explore their data as a network (b), ultimately to investigate relationships in an ecosystem. In this high school history lesson (c), students map global trade networks and learn about the effect of resource location on connections, and connections on power. The full plans can be located via the NiCE website (11).

Our workshop demonstrated a proof-of-principle, clearly eliciting great enthusiasm for network-thinking from the participating teachers. Moreover, it led to the design of several creative and potentially extremely useful module and lesson plans. In future efforts, such plans could be adapted to cater to any grade level. Finally, most participants articulated an intention to take what they had learned back to their home institutions. We will expand on this foundation and initial cohort, and continue to pursue our aim to inspire and empower teachers to take ownership of the network lens for their classrooms.

## Acknowledgements

We gratefully acknowledge conference support from the US Army Research Office: "Network Science and Education", and from the West Point Network Science Center. We thank all of the teachers who came to West Point to participate in this workshop, and shared their passion, experience, and efforts to produce the resources we present here.